\documentclass[twoside]{article}
\usepackage{fleqn,espcrc2,axodraw}

\usepackage{graphicx,psfrag,epsfig,epsf}
\usepackage[figuresright]{rotating}

\newcommand{\AmS}{{\protect\the\textfont2
  A\kern-.1667em\lower.5ex\hbox{M}\kern-.125emS}}

\title{
\begin{flushright}\normalsize
\vspace{-1cm}
{ FERMILAB CONF-04-302-T; IPPP 04/69; PITHA 04/16} \\ 
{ October 2004}
\end{flushright}
Towards pair production near threshold with unstable particle 
effective theory\thanks{Invited talk given by A.S. at the 11th
  International Conference on {\it Quantum Chromodynamics},
  Montpellier, France (5--10th July 2004)}}

\author{
M.~Beneke\address{Institut f\"ur Theoretische Physik E, RWTH Aachen,
D--52056 Aachen, Germany}, 
N.~Kauer\addressmark, 
A.~Signer\address{IPPP, Department of Physics, University of Durham, 
Durham DH1 3LE, England},  
G.~Zanderighi\address{Fermi National Accelerator Laboratory, Batavia, 
        IL 60510-500, USA} }

\begin{document}

\begin{abstract}
\noindent
We illustrate the use of effective theory techniques to describe
processes involving unstable particles close to resonance. First, we
present the main ideas in the context of a scalar resonance in an
Abelian gauge-Yukawa model. We then outline the necessary
modifications to describe $W$-pair production close to threshold in
electron-positron collisions.
\end{abstract}

\maketitle

\section{Introduction}

Processes involving heavy, unstable particles, play an important
role in precision tests of the Standard Model and its hypothetical
extensions. Typically these particles are studied in processes where
they are produced close to resonance. It is thus important to overcome
the difficulties related to the breakdown of ordinary weak-coupling
perturbation theory in the description of such processes. 

As is well known, the singularity of the intermediate propagator can
be avoided if the width, $\Gamma$, of the unstable particle is taken
into account through self-energy resummation. However, this alone 
does not 
guarantee an accurate description of the process. This is partly
reflected by the fact that such a procedure may produce 
gauge-dependent results. Most approaches that have been put forward (for a
summary see~\cite{Grunewald:2000ju}) so far are focused on this
particular aspect of the problem. Recently, we proposed a different
approach~\cite{effth} making use of effective theory methods.  This
allows to perform a systematic expansion in the coupling as well as
in $\Gamma/M$, where $M$ is the mass of the unstable particle. The
main advantage of this method is that it provides us with a
computational scheme for improving the accuracy of
calculations in a systematic way, where gauge invariance is
automatic.

\section{Effective Theory}

At the heart of this method lies the observation that there is a
hierarchy of scales between the mass and the width of the unstable
particle. We exploit this hierarchy by constructing an effective
theory and integrating out the modes that are not needed for the
description of the external state.  A first step to apply these methods
in the context of unstable particles has been made in
\cite{Chapovsky:2001zt} and, as discussed in the next section, the
programme has been carried out explicitly for a toy model in
\cite{effth}.

In technical terms, we identify the relevant modes and use the method
of regions~\cite{Beneke:1998zp} to expand the integrals in
$\Gamma/M$. Together with the standard expansion in the coupling
$\alpha$ we thus achieve a systematic organization of the calculation
in a series in $\alpha$ and $\Gamma/M$.

As a first step in this procedure we integrate out hard momenta $k\sim
M$. The effects of the hard momenta are incorporated into the matching
coefficients, leaving the effective theory without dynamical hard
modes. This is reminiscent and in fact often parallel to similar
developments in other effective theories such as non-relativistic QCD
(NRQCD), heavy quark effective theory (HQET) or soft-collinear
effective theory (SCET). The hard corrections are so called
factorizable corrections, whereas the non-factorizable corrections are
reproduced in the effective theory by the still dynamical
modes~\cite{Chapovsky:2001zt}. The precise nature of these modes
depends to some extent on the underlying theory and even on the
observable in question. In the next sections we will give a 
more detailed discussion, first for a toy model, then for $W$-pair
production near threshold.

\section{An Abelian gauge-Yukawa model}

In order to outline the main steps  in the construction of the  
effective theory (full details can be found in~\cite{effth}) 
%and its use to perform explicit calculations 
we consider a
toy model, consisting of a heavy, scalar field $\phi$ that decays
through a Yukawa interaction into an ``electron'' and
``neutrino''. The electron field, $\psi$, as well as the scalar field
are charged under an Abelian gauge group. The Lagrangian is given by
\begin{eqnarray}
\label{model}
{\cal L} &=& (D_\mu\phi)^\dagger D^\mu\phi - \hat M^2 \phi^\dagger\phi +
 \bar\psi i \!\not\!\!D\psi + \bar\chi i\!\!\not\!\partial\chi
\nonumber \\
&& - \, \frac{1}{4} F^{\mu\nu}F_{\mu\nu}-\frac{1}{2\xi} \,
(\partial_\mu A^\mu)^2
 \nonumber\\
 && + \, y\phi\bar\psi\chi + y^* \phi^\dagger \bar\chi\psi
-\frac{\lambda}{4}(\phi^\dagger \phi)^2+ {\cal L}_{\rm ct}\, ,
\end{eqnarray}
where $\hat{M}$ and ${\cal L}_{\rm ct}$ denote the renormalized mass and the
counterterm Lagrangian respectively.
% and $D_\mu=\partial_\mu-i g A_\mu$. 

For the purpose of illustration we outline the calculation of the
totally inclusive cross section of the process
$\bar{\nu}(q)\, e^-(p) \to X$
%\begin{equation}
%\bar{\nu}(q)\, e^-(p) \to X
%\label{process}
%\end{equation}
as a function of $s\equiv (p+q)^2$ with $s-M^2 \sim M \Gamma \sim
\alpha M^2$, where $\alpha$ denotes collectively the gauge and Yukawa
couplings. We obtain the cross section as the imaginary part of the
forward scattering amplitude ${\cal T}(s)$. The aim is to compute
${\cal T}$ through a systematic expansion in $\alpha$ and
\begin{equation}
\delta \equiv \frac{s-\hat{M}^2}{\hat{M}^2} 
\sim \frac{\Gamma}{\hat{M}} \sim \alpha.
\label{deldef}
\end{equation}
As a first step, we integrate out the hard modes with $k\sim M$. In the
resulting effective theory, the effects of the hard modes are included
in the matching coefficients of the operators. For the heavy scalar,
this procedure is equivalent to the construction of HQET. In fact we
write the momentum of the scalar particle near resonance as
$P^\mu=\hat{M} v^\mu+k^\mu$ where $v^2=1$ and $k\sim M\delta$ and
define the soft field $\phi_v$ (called ``resonant'' in 
\cite{Chapovsky:2001zt}) by removing the rapid spatial
variation $e^{-i\hat{M}\, v\cdot x}$ from the $\phi$-field. Thus, a
soft field has momentum $k \sim M \delta$. We also define the matching
coefficient
%$\Delta \equiv (\bar{s}-\hat{M}^2)/\hat{M}$,
\begin{equation}
\Delta \equiv \frac{\bar{s}-\hat{M}^2}{\hat{M}},
\label{eq:Deltadef}
\end{equation}
where $\bar{s}$ is the complex pole of the propagator. Since $\bar{s}$
and $\hat{M}$ are gauge invariant, $\Delta$ is guaranteed to be gauge
invariant as well. As required for a matching coefficient, $\Delta$ is
given entirely by hard contributions. In fact, expanding the hard part
of the self energy, $\Pi_h(s)$, as
\begin{equation}
\Pi_h(s) = \hat{M}^2 \sum_{k,l} \delta^l\, \Pi^{(k,l)}
\label{eq:Pihexp}
\end{equation}
with $\Pi^{(k,l)}\sim\alpha^k$, we can write the matching coefficient
as
\begin{equation}
\frac{\Delta}{\hat{M}} = \Pi^{(1,0)} +
\left(\Pi^{(2,0)}+\Pi^{(1,1)}\Pi^{(1,0)} \right)
+\ldots .
\label{eq:DeltaPi}
\end{equation}
The explicit form of the matching coefficient depends on the
renormalization scheme. In the pole scheme $\Delta = -i\, \Gamma$. 

The terms of the effective Lagrangian bilinear in $\phi_v$ can be
written as
\begin{eqnarray}
{\cal L}_{\phi\phi} &=&
2 \hat{M} \phi_v^\dagger 
\left( i v\cdot D_s - \frac{\Delta}{2}\right)\phi_v
\label{eq:Lpp} \\
&+& 
2 \hat{M} \phi_v^\dagger 
\left(\frac{(i D_{s\top})^2}{2 \hat{M}} 
   + \frac{\Delta^2}{8 \hat{M}}\right)\phi_v
+\ldots,
\nonumber
\end{eqnarray}
where we defined $D_{s\top}^\mu\equiv D_s^\mu-(v \cdot D_s) v^\mu$. The
interactions of the $\phi_v$-field with soft photons can be
incorporated through the soft covariant derivative $D_s\equiv
\partial-i g A_s$, because the separation into hard and soft parts
respects gauge invariance.

The Lagrangian Eq.(\ref{eq:Lpp}) describes the propagation of an
unstable scalar particle close to resonance. The first line is the
leading term and the inclusion of $\Delta/2$ into the propagator 
corresponds to
self-energy resummation. The terms on the second line are suppressed
by one power of $\delta$ and contribute at NLO. Terms that are
suppressed by further powers of $\delta$ can be included
systematically, if required.

In order to complete the construction of the effective Lagrangian we
have to include kinetic terms for the soft and collinear photons and
fermions, as well as the production and decay vertices for the
unstable particle. In the present Yukawa model the NLO line shape 
is completely described by the addition of the production and 
decay vertices 
\begin{eqnarray}
{\cal L}_{\rm int} &=& C \left[y \phi_v \psi_{n_-}\chi_{n_+} + 
\mbox{h.c.}\right] 
\nonumber\\
&+& \frac{y y^* D}{4\hat{M}^2} (\bar\psi_{n_-}\chi_{n_+})
(\bar\chi_{n_+}\psi_{n_-}),
\end{eqnarray}
where the matching coefficient $C$ must be computed to one loop, 
and $D=1$ at tree level. In this form of writing the effective
Lagrangian we have integrated out the collinear modes, leaving 
only soft and external-collinear modes in the Lagrangian. The latter
describe soft fluctuations around the external light-like momenta. 
Finally, one must compute the scattering process to the one-loop order
in the effective theory. We refer to~\cite{effth} for a 
thorough discussion.

Outside the kinematic region $\delta\sim\alpha$ the effective theory
breaks down. To obtain a consistent description for all values of $s$,
the result of the effective theory has to be matched to an
off-resonance calculation in the full theory.

\section{$W$-pair production close to threshold}

The toy model presented above is in many ways much simpler
than the Standard Model, but it contains all the relevant features 
of resonant particle
production. Applying the same techniques to a realistic process
may result in additional technical complications, but no new
conceptional difficulties arise.

An important application is $W$-pair
production close to threshold at an electron-positron collider. This
process is crucial for the precise determination of the $W$ mass and
has been thoroughly studied away from threshold. 
In particular, the one-loop electroweak
corrections have been computed in the double pole approximation
(DPA)~\cite{ww-calc}. Here we focus on obtaining results that are 
valid near threshold, where the DPA
is supposed to break down.

As mentioned before, the first step is to integrate out the hard
modes. For pair production close to threshold, the $W$ bosons 
are then described by a non-relativistic Lagrangian,
analogous to NRQCD and the complications due to the Coulomb
singularity can be addressed using standard methods. We are thus left
with dynamical degrees of freedom~\cite{Beneke:1998zp} familiar from
NRQCD~\cite{nrqcd}. We particularly emphasize the presence of soft and
potential gauge bosons. Working in the center of mass frame, the
typical momentum of a $W$ with velocity $v$ is $|\vec{k}| \sim M_W\,
v$ and the typical (non-relativistic) energy is $E\sim M_W\,
v^2$. This corresponds to a potential mode. For soft modes, the energy
and momentum scale as $M_W\, v$. The initial state fermions and the
decay products are described by collinear modes familiar from
SCET~\cite{scet}. In fact we will need several collinear modes, one
for each direction present in the process under consideration.

To be specific, let us consider the process
\begin{equation}
e^-(p_1)\, e^+(p_2) \to 
\mu^-(l_1)\, \bar{\nu}_\mu(l_2)\, u(l_3)\, \bar{d}(l_4)
\label{wwprocess}
\end{equation}
with $(l_1+l_2)^2-M_W^2 \sim (l_3+l_4)^2-M_W^2 \sim M_W^2 v^2
\sim M_W\Gamma_W$. By assumption $\sqrt{s}/2-M_W \sim M_W\Gamma_W$.
%More
%precisely, we parametrize the momenta as follows: $p_1 = (\sqrt{s}/2,
%\vec{p})$, $p_2 = (\sqrt{s}/2, -\vec{p})$, $k_1 = P/2+k$, $k_2 = P/2
%-k$, where $|\vec{p}\,| = \sqrt{s}/2$ and, by assumption
%$\sqrt{s}/2-M_W\sim k^0\sim M_W\, v^2 \sim M_W\, \alpha_{ew}$ and
%$|\vec{k}|\sim M_W\, v$. We also made use of the definitions $P\equiv
%p_1+p_2$, $k_1 \equiv l_1+l_2$ and $k_2 \equiv l_3+l_4$.

At leading order, the description of the process can be split into
three parts. First, the $W$-pair is produced, then it propagates, and
finally the $W$ bosons decay. At higher orders, this simple picture is
complicated by the exchange of Coulomb photons, of photons that
connect the various stages and/or by the presence of additional
photons in the final state. Furthermore, single-resonant and
non-resonant diagrams have to be taken into account. We count
$\alpha_s\sim v \sim\sqrt{\alpha_{ew}}$ and by next-to-leading order
(NLO) contributions we understand the ${\cal O}(\alpha_{ew}) \sim{\cal
O}(v^2)\sim {\cal O}(\alpha_{s}^2)$ corrections. In this article we
present all contributions that are needed at $\sqrt{\rm N}$LO, that
is, all corrections of order $v\sim\alpha_s$.

Starting with the construction of the non-relativistic Lagrangian we
first remark that the gauge
dependence of the vector-boson propagator is not an issue. 
Indeed, expanding the
momentum of a non-relativistic $W$ as $k^\mu = M_W\, v^\mu + q^\mu$,
where $v^\mu\equiv(1,\vec{0})$, and $q^\mu=(q^0,\vec{q})$ and taking
into account $q^0\sim M_W\, v^2$ and $|\vec{q}\, |\sim M_W\, v$ the
propagator in a general $R_\xi$ gauge upon expansion becomes
\begin{eqnarray}
&& \hspace*{-0.5cm} \frac{-i}{k^2-M_W^2} 
\left(g^{\mu\nu} - (1-\xi)\frac{k^\mu k^\nu}{k^2-\xi M_W^2} \right)
\nonumber \\
&\to&
\frac{-i (g^{\mu\nu}-v^\mu v^\nu)}{k^2-M_W^2}
\to \frac{i\, \delta^{ij}}{2 M_W\, q^0 - \vec{q}^{\,2}}.
\label{eq:Wprop}
\end{eqnarray}
The propagator scales as $v^{-2}$, is gauge independent and describes
three polarization states of a non-relativistic particle. 
This is consistent with the fact that the
degrees of freedom with mass $\sqrt{\xi} M_W$ have been integrated
out. (In Feynman gauge the propagator is given by $-i\, g^{\mu\nu}/(2
M_W\, q^0 - \vec{q}^{\, 2})$. The effects of the unphysical scalar
polarization are cancelled by the pseudo-Goldstone field, which has
mass $M_W$ and is still present in the effective theory.)  Including
the decay width we obtain for the non-relativistic Lagrangian up to
NLO
\begin{eqnarray}
{\cal L}_{nr} &=& \sum_\mp \Omega_\mp^{* i} \left(
i D^0 + \frac{\vec{D}^2}{2 \hat{M}_W} - \frac{\Delta}{2} \right) 
\Omega_\mp^i
\nonumber \\
&+&  \sum_\mp \Omega_\mp^{* i}\, 
\frac{(\vec{D}^2-\hat{M}_W \Delta)^2}{8 \hat{M}_W^3}\, 
\Omega_\mp^i,
\label{LNR}
\end{eqnarray}
where $\Omega_\pm$ denote the non-relativistic vector fields with mass
dimension 3/2 for the $W^\pm$ bosons.  The matching coefficient
$\Delta\equiv (\bar{s}-M_W^2)/M_W$ is defined as before. Its leading part,
$\Delta^{(1)}\sim M_W \alpha_{ew}$ scales as $D^0$ and $\vec{D}^2$ and,
thus, has to be included in the leading order Lagrangian. 
%This
%corresponds to the resummation of the self-energy
%insertions. 
Accordingly, the propagator of a $\Omega_\pm$ with energy
$E$ and momentum $\vec{k}$ is given by
\begin{equation}
\frac{i\, \delta^{ij}}{\left(E - \frac{\vec{k}^2}{2 M_W}
  -\frac{\Delta^{(1)}}{2}\right)}.
\label{OmegaProp}
\end{equation}
Higher order corrections to $\Delta$ can either be resummed,
i.e. included in the propagator, or included perturbatively as 
interactions. At $\sqrt{\rm N}$LO we need $\Delta^{(3/2)}$, the ${\cal
O}(\alpha_{ew} \alpha_s)$ corrections to $\Delta$. Higher order
corrections are at least NLO as are the terms in the second line
of Eq.(\ref{LNR}).

The production of a $W^- W^+$ pair is described by effective vertices
to be added to ${\cal L}_{nr}$. In order to obtain the corresponding
operators and their matching coefficients, we have to compute the
amputated, renormalized on-shell Green function for $e^+e^-\to W^+
W^-$ to the desired order in ordinary weak-coupling perturbation
theory. At leading order, only the helicity configuration $e^-_L
e^+_R$ contributes and the corresponding operator reads
\begin{equation}
{\cal L}_p^{(0)} = \frac{\pi\alpha_{ew}}{M_W^2} 
\left(\bar{e}_L \gamma^{[i} n^{j]} e_L \right) 
\left(\Omega_-^{*i} \Omega_+^{*j}\right) ,
\label{LPlead}
\end{equation}
where $e_L$ ($\bar e_L$) is an external-collinear field~\cite{effth} 
with large momentum in the $\vec{n}$ ($-\vec{n}$) 
direction and we introduced the notation
$a^{[i} b^{j]}\equiv a^i b^j + a^j b^i$. Including terms that are
suppressed by one power of $v$ we get additional operators
\arraycolsep0.1cm
\begin{eqnarray}
{\cal L}_p^{(1/2)} &=& \frac{c_1}{M_W^3}
\left(\bar{e}_L \gamma^{j} e_L \right) 
\left(\Omega_-^{*i} (-i) D^j \Omega_+^{*i}\right)
\label{LPhalf} \\
&+& \frac{c_2}{M_W^3}
\left(\bar{e}_L \gamma^{[i} e_L \right) 
\left(\Omega_-^{*i} (-i) D^{j]} \Omega_+^{*j}\right) 
\nonumber \\
&+& \frac{c_3}{M_W^3}
\left(\bar{e}_L \gamma^{[i} n^{j]} n^l e_L \right) 
\left(\Omega_-^{*i} (-i) D^l \Omega_+^{*j}\right) 
\nonumber \\
&+& \frac{c_4}{M_W^3}
\left(\bar{e}_L \gamma^j\gamma^l\gamma^i e_L \right) 
\left(\Omega_-^{*i} (-i)D^l \Omega_+^{*j}\right)
\nonumber
\end{eqnarray}
with the matching coefficients
\begin{eqnarray}
c_1 &=& \pi\alpha_{ew}
\frac{M_Z^2 \sin^2\theta_w -2 M_W^2}{4 M_W^2-M_Z^2}
\label{eq:c1} \\
c_2 &=& \pi\alpha_{ew} 
\frac{M_Z^2 (1-2 \sin^2\theta_w)}{4 M_W^2-M_Z^2}
\label{eq:c2} \\
c_3 &=& 2 \pi\alpha_{ew}  \phantom{\frac{1}{2}}
\label{eq:c3} \\
c_4 &=& \pi\alpha_{ew}
\label{eq:c4}
\end{eqnarray}
All derivatives on the potential $W^\pm$ fields 
in Eq.(\ref{LPhalf}) scale as $D^i\sim
M_W v$.  We also remark that the $e^-_R\, e^+_L$ helicity configuration
does not vanish at $\sqrt{\rm N}$LO.  Consequently, there
are additional operators in ${\cal L}_p^{(1/2)}$ involving
$\bar{e}_R$ and $e_R$.

The decay of the vector bosons can be described in a similar way as 
the production, namely through decay vertices. The operators
with their matching coefficients are obtained through matching of the
corresponding on-shell Green function. For the leptonic decay, at
$\sqrt{\rm N}$LO there are no loop corrections to be taken into
account, whereas for the hadronic decay, there are corrections of the
order $\alpha_s\sim v$. When the calculation is 
completely inclusive on the hadronic
decay products, these corrections, together with the corrections from
final state emission of an additional gluon, are taken into account by
the hadronic part of the decay width. For the decay operators that
are relevant to the process Eq.(\ref{wwprocess}) we obtain
\begin{equation}
{\cal L}^{(0)}_d = 
-\frac{g_{ew}}{2\sqrt{M_W}}\left( \Omega_-^i \bar{\mu}_L \gamma^i \nu_L
+ \Omega_+^i \bar{u}_L \gamma^i d_L \right) .
\label{LDlead}
\end{equation}
Kinematic corrections to ${\cal L}^{(0)}_d$ are suppressed by at least
two powers of $v$.

%%%%%%%%%%%%%%%%%%%%%%%%%%%%%%%%%%%%%%%%%%%%%%%%%%%%%%%%%%%%%%%%%%%
\begin{figure}[h]
%\psfrag{S}{$S_{\pi\pi}$}
%\psfrag{etabar}{\hspace*{0.5cm}$\bar\eta$}
\begin{center}
\begin{picture}(150,80)(0,0)
\ArrowLine(40,40)(10,10)
\ArrowLine(10,70)(40,40)
\Text(5,10)[r]{$p_2$}
\Text(5,70)[r]{$p_1$}
\GCirc(40,40){2}{0.1}
\Photon(38,40)(100,67){-2}{6.5}
\Photon(38,40)(100,13){2}{6.5}
\Text(70,65)[c]{$k_1$}
\Text(70,15)[c]{$k_2$}
\GCirc(100,65){2}{0.1}
\GCirc(100,15){2}{0.1}
\ArrowLine(100,65)(136,80)
\ArrowLine(136,50)(100,65)
\ArrowLine(100,15)(136,0)
\ArrowLine(136,30)(100,15)
\Text(140,80)[l]{$l_1$}
\Text(140,50)[l]{$l_2$}
\Text(140,0)[l]{$l_3$}
\Text(140,30)[l]{$l_4$}
\end{picture}
\end{center}
\vspace{-1.cm} {\caption{\it Leading order Feynman diagram in the
effective theory.
\label{fig:LO}}}
\vspace{-0.4cm}
\end{figure}
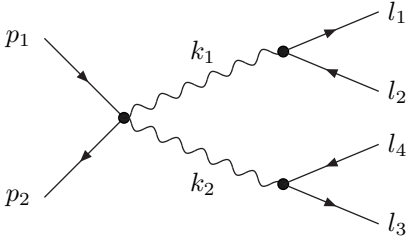
%%%%%%%%%%%%%%%%%%%%%%%%%%%%%%%%%%%%%%%%%%%%%%%%%%%%%%%%

We are now in a position to compute the leading order amplitude within
the effective theory. The corresponding diagram is depicted in
Figure~\ref{fig:LO}. For the vertices we have to take the leading
order production and decay vertices, Eqs.(\ref{LPlead}) and
(\ref{LDlead}). The propagators of the intermediate vector bosons are
given by Eq.(\ref{OmegaProp}). Putting everything together we obtain
\begin{eqnarray}
{\cal A}^{(0)} &=& \frac{i \alpha_{ew}^2 \pi^2}{M_W^3}\, 
\langle p_2-|n^{[i} \gamma^{j]} |p_1-\rangle \times
\label{eq:Alead} \\
&& \hspace*{-1.5cm}
\frac{
\langle l_1-|\gamma^i|l_2-\rangle
\langle l_3-|\gamma^j|l_4-\rangle}{
\left(E_1 - \frac{(\vec{l}_1+\vec{l}_2)^2}{2 M_W} 
    - \frac{\Delta^{(1)}}{2}\right)
\left(E_2 - \frac{(\vec{l}_3+\vec{l}_4)^2}{2 M_W} 
    - \frac{\Delta^{(1)}}{2}\right)} ,
\nonumber
\end{eqnarray}
where $E_1\equiv l_1^0+l_2^0-M_W$, $E_2\equiv
l_3^0+l_4^0-M_W$ and we used standard helicity notation.

Turning to the calculation of the $\sqrt{\rm N}$LO amplitude, we also
have to consider the helicity configuration $e_R^-\, e_L^+$. However,
the corresponding calculation is analogous to the calculation of
${\cal A}^{(0)}$, barring the replacement of the production
vertex. Let us, therefore, focus on the corrections needed in the case
of $e_L^-\, e_R^+$.

For all stages, production, propagation and decay we have to
include corrections. In the case of the production stage, we have to
include diagrams as shown in Figure~\ref{fig:LO} with the production
vertex due to operators given in Eq.({\ref{LPhalf}) rather than
Eq.({\ref{LPlead}). For the propagation stage, we need to include the
${\cal O}(\alpha_{ew} \alpha_s)$ corrections to $\Delta$. If resummed,
this results in a change in the propagator $\Delta^{(1)}\to
\Delta^{(1)} + \Delta^{(3/2)}$. Finally, regarding the decay stage, we
mention again that the ${\cal O}(\alpha_s)$ corrections 
to the decay
vertex are not explicitly needed in a completely inclusive
calculation. 
%as long as we are completely
%inclusive in the hadronic decay products.

Apart from these trivial corrections, there is a further contribution at
$\sqrt{\rm N}$LO, namely the exchange of a single Coulomb (potential) photon,
shown in Figure~\ref{fig:Coulomb}. This contribution is suppressed by
$\alpha/v \sim v$ relative to the leading order amplitude. Thus,
contrary to top pair production close to threshold, these
contributions can be treated perturbatively and need not be
summed. This is the only correction at $\sqrt{\rm N}$LO that is not
due to hard modes.

%%%%%%%%%%%%%%%%%%%%%%%%%%%%%%%%%%%%%%%%%%%%%%%%%%%%%%%%%%%%%%%%%%%
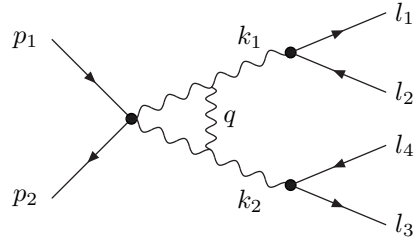
\begin{figure}[h]
%\psfrag{S}{$S_{\pi\pi}$}
%\psfrag{etabar}{\hspace*{0.5cm}$\bar\eta$}
\begin{center}
\begin{picture}(150,80)(0,0)
\ArrowLine(40,40)(10,10)
\ArrowLine(10,70)(40,40)
\Text(5,10)[r]{$p_2$}
\Text(5,70)[r]{$p_1$}
\GCirc(40,40){2}{0.1}
\Photon(38,40)(100,67){-2}{6.5}
\Photon(38,40)(100,13){2}{6.5}
\Photon(70,52)(70,28.3){2}{4}
\Text(75,40)[l]{$q$}
\Text(85,71)[c]{$k_1$}
\Text(85,9)[c]{$k_2$}
\GCirc(100,65){2}{0.1}
\GCirc(100,15){2}{0.1}
\ArrowLine(100,65)(136,80)
\ArrowLine(136,50)(100,65)
\ArrowLine(100,15)(136,0)
\ArrowLine(136,30)(100,15)
\Text(140,80)[l]{$l_1$}
\Text(140,50)[l]{$l_2$}
\Text(140,0)[l]{$l_3$}
\Text(140,30)[l]{$l_4$}
\end{picture}
\end{center}
\vspace{-1.cm} {\caption{\it The exchange of a single Coulomb photon
    is suppressed by $\alpha/v$ relative to the leading order
    amplitude and, thus, contributes at $\sqrt{N}$LO.
\label{fig:Coulomb}}}
\vspace{-0.4cm}
\end{figure}
%%%%%%%%%%%%%%%%%%%%%%%%%%%%%%%%%%%%%%%%%%%%%%%%%%%%%%%%

The momentum of the potential Coulomb photon scales as $q^0\sim M_W \,
v^2$, $|\vec{q}\,|\sim M_W\, v$. Thus the propagator is given by
$i/\vec{q}^{\, 2}$. Reading off the Feynman rules of the $WW\gamma$
vertex of Eq.(\ref{LNR}), we obtain for the amplitude due to single
photon exchange
\begin{eqnarray}
{\cal A}^{(1/2,c)} &=& -i\, (4\pi\alpha)\, {\cal A}^{(0)} \times
\label{eq:Acoul} \\
&& \hspace*{-1.5cm}
\int \frac{d^D q}{(2\pi)^D}\, \frac{1}{\vec{q}^{\,2} }\,
\frac{1}{\left(E_1 - q^0 - \frac{(\vec{k}_1-\vec{q})^2}{2 M_W} 
      - \frac{\Delta^{(1)}}{2} + i\epsilon \right) } 
\nonumber \\
&& \hspace*{-0.2cm}
\frac{1}{\left(E_2 + q^0 - \frac{(\vec{k}_2+\vec{q})^2}{2 M_W} 
      - \frac{\Delta^{(1)}}{2} + i\epsilon\right) } ,
\nonumber
\end{eqnarray}
where $\vec{k}\equiv\vec{k}_1=\vec{l}_1+\vec{l}_2$ and
$\vec{k}_2=\vec{l}_3+\vec{l}_4=-\vec{k}$.  After performing the $q^0$
contour integral, the $\vec{q}$ integration is straightforward and we
obtain
\begin{eqnarray}
{\cal A}^{(1/2,c)} &=& {\cal A}^{(0)} \, 
\frac{\alpha\, M_W}{|\vec{k}|} \times
\label{eq:AcoulF} \\
&& \hspace*{-1cm}
\arctan\frac{|\vec{k}|}
  {\sqrt{M_W(\Delta^{(1)}-E_1-E_2) - i\epsilon}},
\nonumber 
\end{eqnarray}
in agreement with~\cite{Fadin:1993kg}. 

We note that it is possible to
resum the contributions due to multiple potential photon exchange and
derive a Green function for the $W$ pair. This is described most
naturally within the context of a potential non-relativistic Lagrangian,
analogous to potential non-relativistic QED~\cite{Pineda:1998kn}.

\section{Conclusions}

The fundamental reason for the breakdown of ordinary weak-coupling
perturbation theory for processes involving resonant unstable
particles is the appearance of a second small scale. We described a
method that overcomes these problems using an effective theory
approach and applied this first to a toy model involving an unstable
scalar particle and then to $W$-pair production near threshold.

The applications we presented are technically rather simple. In the
case of $W$-pair production this is because we have considered only
the ${\cal O}(v)$ corrections. However, we would like to stress once
more, that including higher-order corrections introduces only
technical problems and no new conceptual problems arise.

In this article, we have restricted ourselves to a totally inclusive
cross section. More complicated final state kinematics
requires in general the introduction of additional modes in the
effective theory. Also, phase-space integrals have to be
expanded. This is done most conveniently by working directly with
cut-diagrams, rather than amplitudes, because this facilitates the use
of the same methods as for loop integrals. Thus, the method presented
here is not restricted to a small number of special cases. It provides
us with a consistent systematic computational scheme that can be
applied to a wide variety of processes.

\section*{Acknowledgements}

The work of M.B. and N.K. is supported in part by the DFG
Sonderforschungsbereich/Transregio 9 ``Computergest\"utzte
Theoretische Teilchenphysik''

\end{document}